\documentclass[conference]{IEEEtran}
\IEEEoverridecommandlockouts
\usepackage{cite}
\usepackage{amsmath,amssymb,amsfonts}
\usepackage{algorithmic}
\usepackage{graphicx}
\usepackage{textcomp}
\usepackage{xcolor}
\usepackage{verbatim}
\def\BibTeX{{\rm B\kern-.05em{\sc i\kern-.025em b}\kern-.08em
    T\kern-.1667em\lower.7ex\hbox{E}\kern-.125emX}}

\begin{document}

\title{Deign of an Internet of Things System for Smart Hospitals\\
}

\author{\IEEEauthorblockN{1\textsuperscript{st} Jichao Leng}
\IEEEauthorblockA{\textit{dept. name of organization (of Aff.)} \\
\textit{name of organization (of Aff.)}\\
City, Country \\
email address or ORCID}
\and
\IEEEauthorblockN{2\textsuperscript{nd} Given Name Surname}
\IEEEauthorblockA{\textit{dept. name of organization (of Aff.)} \\
\textit{name of organization (of Aff.)}\\
City, Country \\
email address or ORCID}
\and
\IEEEauthorblockN{3\textsuperscript{rd} Given Name Surname}
\IEEEauthorblockA{\textit{dept. name of organization (of Aff.)} \\
\textit{name of organization (of Aff.)}\\
City, Country \\
email address or ORCID}
\and
}

\author{Jichao Leng, Xucun Yan, Zihuai Lin\\
	School of Electrical and Information Engineering, The University of 
	Sydney, Australia\\
	Emails:	\{jichao.leng,\ xucun.yan,\ zihuai.lin\}@sydney.edu.au. 
}

\maketitle

\begin{abstract}
With the rapid development of smart devices and the Internet of Things (IoT) technology, some traditional scenarios are exploring new possibilities. Especially in the field of healthcare, the mixed and large amount of people, the complex and professional data, and the strict environmental requirements on some medical scenes and equipment, all put forward extremely high requirements on the hospital management. Therefore, an efficient and secure Internet of things system is greatly necessary.
This paper develops an IoT system that could be deployed in hospitals for various applications. This system supports LoRa, Wi-Fi and other data collection methods, uploads the data to the cloud platform for further processing through secure connection, and finally feeds back to users in real-time through the user interface. This system supports accurate indoor positioning based on UWB, ECG signal detection, environmental monitoring, people flow statistics and other functions.
\end{abstract}
\smallskip

\section{Introduction}
The use of the Internet of Things is growing steadily over the years. By 2020, there will be more than 20 billion interconnected IoT devices, and the market size may reach 1.5 trillion dollars. It is expected that each person will have an average of four connected devices. At present, the Internet of Things is one of the main promoters of technological innovation and one of the areas with greater potential for social and economic transformation. 

All involved stakeholders, from technicians to developers, companies and users, face several challenges that remain to be resolved. Experience in distributed systems, networks, mobile and pervasive computing, context awareness and WSN can be considered a good starting point for finding appropriate solutions such as interoperability, openness, security, scalability and troubleshooting problem\cite{1}. The key to the IoT system project is the timely collection and processing of data from different sources. Cross-region resource allocation and big data analysis processing are made possible by building an IoT network in the scene.

A hospital is a place with large people flow and complicated personnel. The traditional hospital management method is difficult to carry out unified management and resource allocation for the entire hospital area, which leads to a lot of redundant manpower and resources wasted in this aspect. For that, the hospital IoT system can provide efficient and refined management for all kinds of people and regions in the hospital\cite{2}. 

Our work will be used as a hospital scene to improve the operational efficiency and management capabilities of the entire scenario by setting up a complete and multi-functional IoT system. For example, for environmental monitoring of designated areas, such as drug storage rooms or operating rooms, where temperature and humidity are critical, users can view the real-time status of all these areas and the corresponding alarm thresholds through the platform, and even adjust them automatically as needed. For the indoor positioning function, it provides an efficient resource and personnel management method for managers. 

When an emergency occurs, the administrator can directly view the real-time location of a specific doctor through the platform, or directly inform the nearest doctor of the event. For the people flow statistics function, users can view the flow of people and congestion in various areas of the hospital in real-time and adjust the resources of each department according to the feedback to improve efficiency. At the same time, patients will be provided with wearable real-time ECG and posture detection devices so that hospital staff can respond to any abnormal situation with location information.

\begin{figure}[htbp]
\centering{\includegraphics[width = 87 mm]{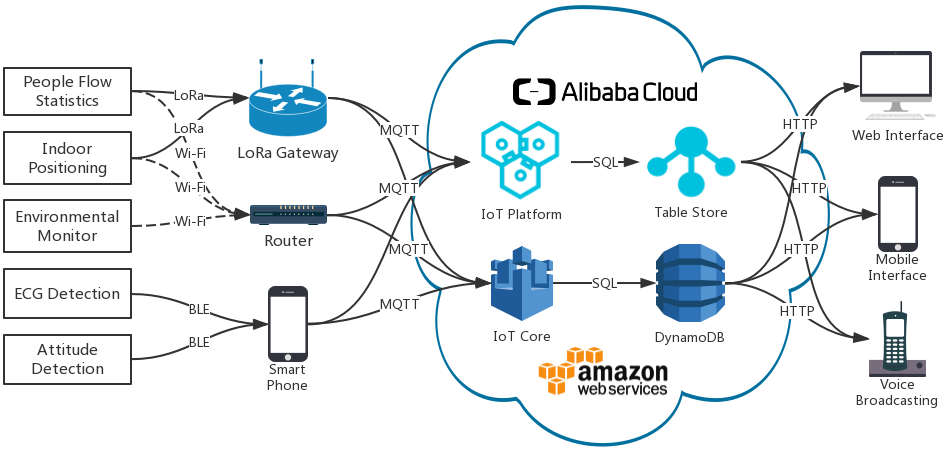}}
\caption{System Framework.}
\label{fig.fram}
\end{figure}

In this paper, we first introduce the whole system framework with dividing it into five layers and introduce the functions of each layer in Section II. The framework of our system is shown in Fig.\ref{fig.fram}. In Section III, the functions and hardware deployment of each module in the system are explained in detail. In Section IV, we design a demonstration to test the functions and the interfaces of this system. In Conclusion, the direction of the next stage of work is put forward. 

\section{System Architecture}
\begin{figure}[htbp]
\centering{\includegraphics[width = 87 mm]{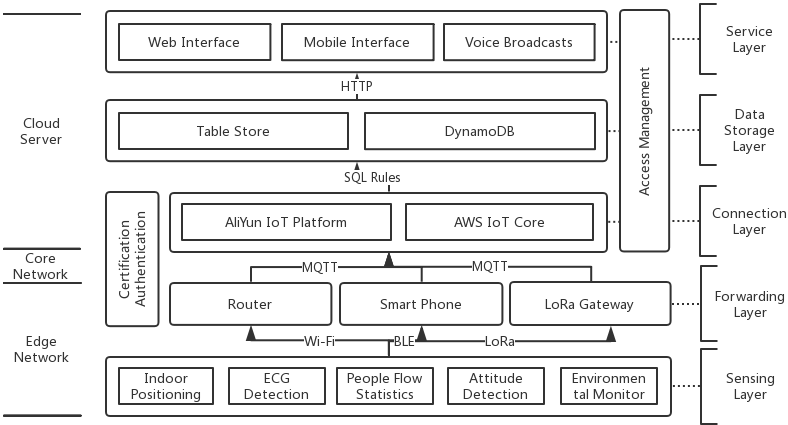}}
\caption{Five-layer System Structure.}
\label{fig.stru}
\end{figure}

Our system implements a complete IoT system \cite{IoT_FD}, including sensor data acquisition, edge network formation and gateway data forwarding, cloud device authentication and data processing, database storage, data recovery and visualization. Finally, all collected data is fed back to the user in an intuitive graphical interface. As shown in Fig.\ref{fig.stru}, the system is divided into five layers according to functions: Sensing Layer, Forwarding Layer, Connection Layer, Data Storage Layer and Service Layer. The sensing layer contains functional devices based on a variety of wireless protocols for the IoT. Heterogeneous data from different devices will be processed in a unified form in the forwarding layer, and the data will be sent to the cloud through establishing a secure connection with the connection layer and stored in the database in the data storage layer. This data will be fed back to the user through the user interface in the service layer.

\subsection{Sensing Layer}
In the sensing layer, data acquisition devices from various functions carry different types of sensors, which can produce different types of data containing various information and sometimes even affecting each other. This system has established two edge networks based on LoRa and Wi-Fi respectively and has deployed several functional applications based on different wireless protocols, such as indoor positioning \cite{pos1,pos2} based on UWB and BLE, ECG monitoring device \cite{IREALCARE1,IREALCARE2,IREALCARE3,IREALCARE4,IREALCARE5} based on Bluetooth, and people flow statistics based on ultrasonic and Wi-Fi. In the future work, radio sensing techniques, e.g., \cite{GI1,GI2,GI3,GI4}, can also be used in our system.

\subsection{Forwarding Layer}
In the forwarding layer, the original data from the sensor layer will be re-encapsulated, the identifier and timestamp will be added according to the source of the data, and the data will be uploaded to the cloud through the reliable connection established with the connection layer. In this system, a smart IoT gateway is used to process local data and communicate with the cloud through MQTT protocol.

\subsection{Connection Layer}
In the connection layer, the system establishes the corresponding cloud map for each sensor layer device and uses the corresponding interface to receive the message uploaded from the forwarding layer. The received messages are filtered and collated and then stored in a specific format in the appropriate database for subsequent calls. In this system, the AWS and Alibaba Cloud are used as the cloud platform respectively.

\subsection{Data Storage Layer}
In the data storage layer, table storage is used as the database form. A table store is a fully managed NoSQL database that stores data as a sorted map. The data forwarded by the connection layer is stored in the corresponding form according to the source function represented by the identifier and sorted by timestamp. The TTL of the data is adjusted according to the functional requirements.

\subsection{Service Layer}
The service layer provides users with the functions of data visualization and voice broadcasting, thus allowing users to get real-time data feedback. Users can access the user interface provided by the system through web pages or mobile APP, and the data source comes from the corresponding database of the data storage layer. As an auxiliary system, voice broadcast can meet the needs of special occasions.

\section{System Deployment}
In this section, the software methods and hardware devices involved in the functional realization of each process of the system will be introduced respectively. Especially in the sensing functions part, several important function modules in the system will be introduced in detail.

\subsection{Sensing Functions}
At present, several important function modules are realized in the sensor layer of the system: 

1) An indoor positioning functions combining UWB and BLE: this function can locate personnel, equipment or other materials in the hospital in real-time, and feedback the location information on the user interface for easy search and management. It is accurate to the centimetre level and has an update delay of less than one second. 

2) A wearable ECG signal detection function: this device \cite{IREALCARE1,IREALCARE2,IREALCARE3,IREALCARE4,IREALCARE5} is carried by patients and will monitor and record the user's ECG signal waveform in real-time and upload it to the cloud for professional analysis. 

3) An ultrasonic people flow statistics function, assisted by Wi-Fi probe: this function can monitor the flow of people in the area in real-time. For the scene of huge people amount in the hospital, the monitoring of people traffic flow in different areas can allow the manager to take some measures in time to avoid potential risks.

4) A wearable function for monitoring falls of the elderly: this device is carried by the patient. When the sensor of the device detects abnormal changes in acceleration in some directions (for example, the patient falls), the cloud will display the alarm information. 

5) An environmental monitoring functions: the corresponding sensors are used to detect the temperature, humidity, noise decibel, magnetic field intensity and other environmental indicators in the target environment, and record the changing trend of various indicators in the hospital. 

\paragraph{Indoor Positioning} 
In this system, a set of indoor positioning system combined with UWB and BLE is adopted. The UWB positioning can be used to achieve more accurate positioning in areas with dense departments and complex environment, while for other areas, the BLE positioning with lower cost can be adopted. The expanded board X-NUCLEO-IDB05A1 is used for BLE positioning. For the BLE position, the target location at the moment will be confirmed through the RSSI of the measurement point target and the location nodes, and the data will be uploaded through LoRa by SX1272 expanded board. The UWB positioning uses DW1000FOLLOWER to determine the distance between the target and the positioning nodes by measuring the TDOA, which is calculated as:
\begin{equation*}
D=\frac{\left ( T_{round1}\times T_{round2}-T_{reply1}\times T_{reply2} \right )}{\left ( T_{round1}+ T_{round2}+T_{reply1}+ T_{reply2} \right )}\times c
\end{equation*}
where $T_{round}$ is the total flight time, $T_{reply}$ is the processing time. In the region, three UWB anchor points will be used to obtain the distance between the target and the three anchor points respectively, and the coordinate of the target will be determined by the trilateral positioning method. At present, the least square method is used in this system to deal with the errors generated in the positioning process. In future work, CNN will be used to reduce positioning errors.

\begin{figure}[htbp]
\begin{minipage}[t]{0.5\linewidth}
\centering 
\includegraphics[width=44 mm]{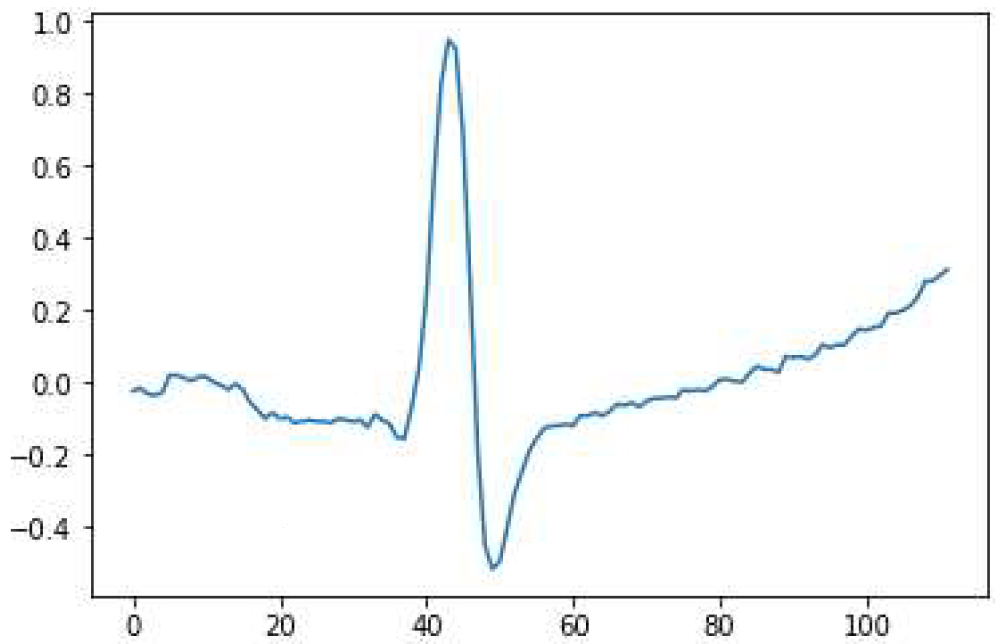}
\\
(a) Original Signal.
\\
\end{minipage}%
\begin{minipage}[t]{0.5\linewidth} 
\centering 
\includegraphics[width=44 mm]{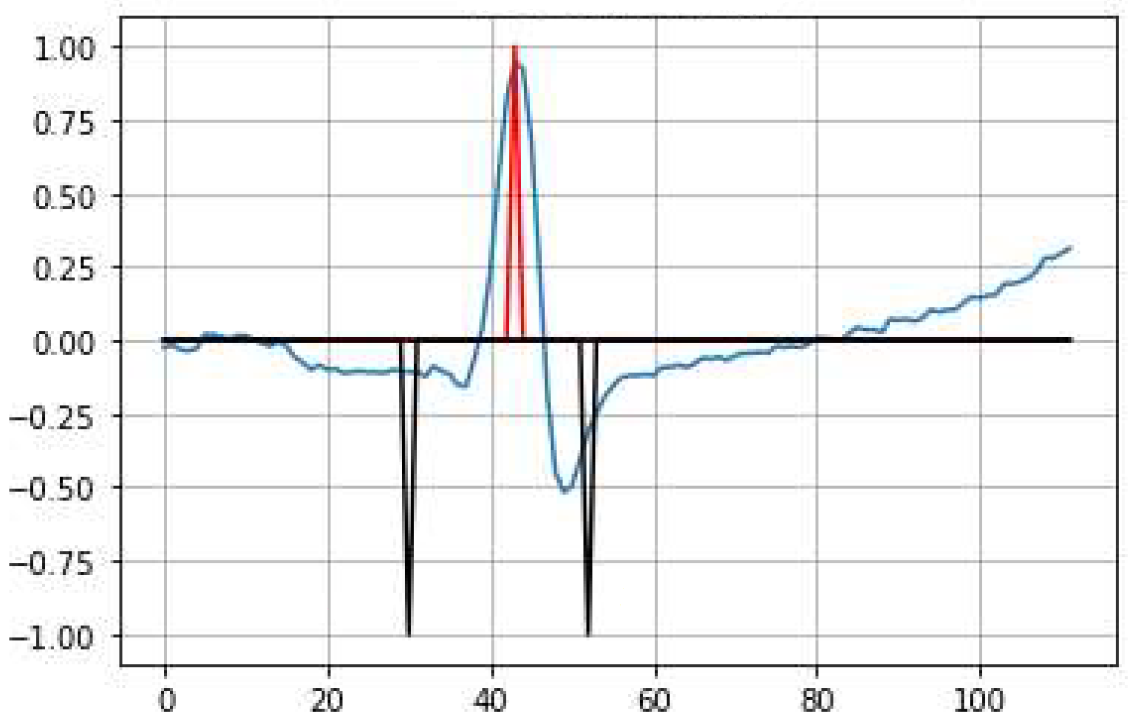} 
\\
(b) QRS Complex Analysis.
\\
\end{minipage} 
\label{fig.ecg} 
\caption{ECG Signal Detection.} 
\end{figure}

\paragraph{ECG Detection}
In this system, a wearable ECG device Holter monitor IREALCARE is used to detect ECG signals of patients. The sensor records the real-time ECG signal curve and sends it to the smartphone via Bluetooth, which is then uploaded to the cloud for further analysis. The original data collected by the sensor is shown in Fig.\ref{fig.ecg}(a). The QRS complex will be found through waveform slope analysis like in Fig.\ref{fig.ecg}(b), and the selected signal will be fed into a self-built CNN model to analyze the health condition of users. 

\paragraph{People Flow Statistics} 
The people flow statistics function includes two modules to realize the people flow management in the scene. The proximity sensors are deployed in each gate to calculate the number of people entering and exiting through the gate and the total number. Two adjacent VL53L0X sensors are used to judge the target's direction of travel, and its data is sent to the gateway for further processing by LoRa SX1272 expanded board. Another is to estimate the number of people in the target area by counting the number of MAC addresses of mobile devices through Datasky DS006 Wi-Fi probe, and the result is sent to the cloud through HTTP protocol.

\begin{figure}[htbp]
\begin{minipage}[t]{0.5\linewidth}
\centering
\includegraphics[width=44 mm]{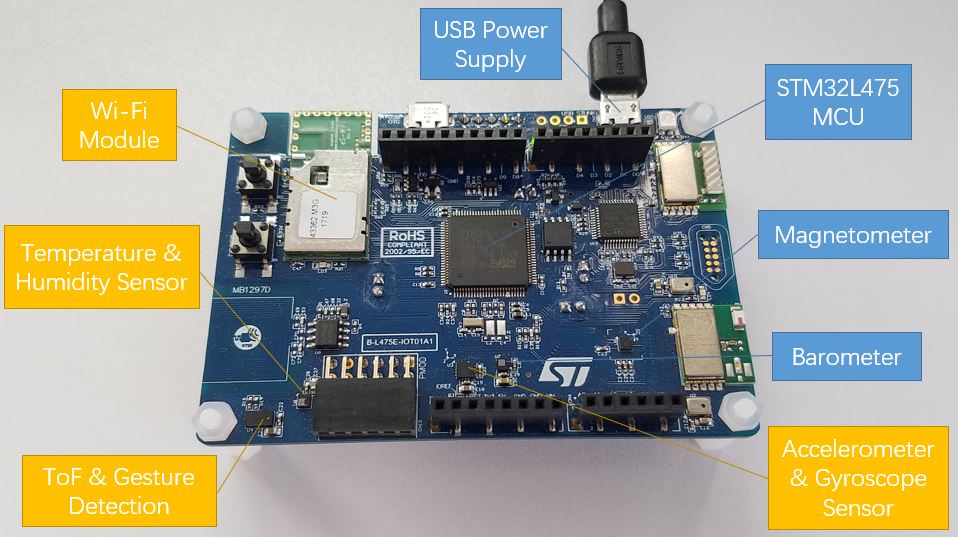} 
\\
(a) B-L475E-IOT01A.
\\
\end{minipage}%
\begin{minipage}[t]{0.5\linewidth} 
\centering 
\includegraphics[width=44 mm]{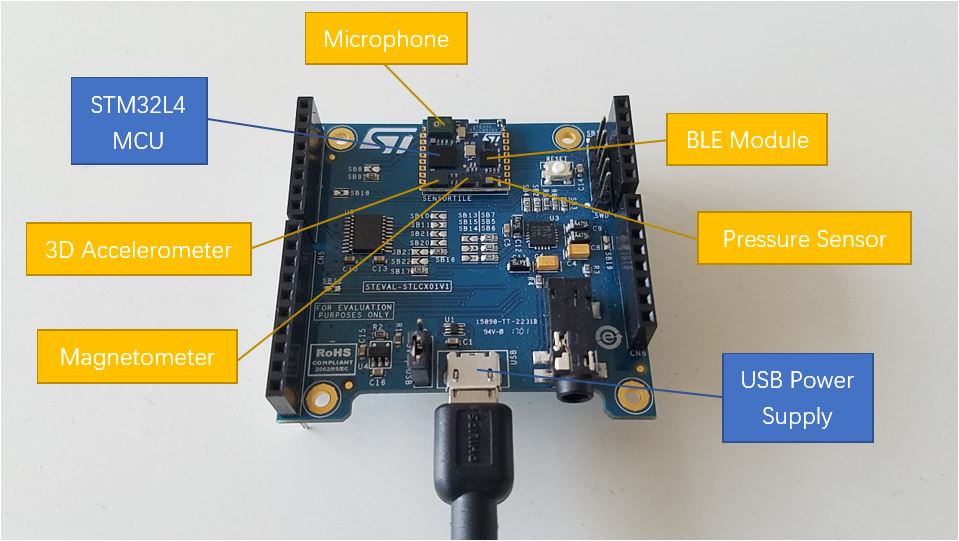}
\\
(b) STEVAL-STLKT01V1.
\\
\end{minipage} 
\caption{Sensing Device.}
\label{fig.st}
\end{figure}

\paragraph{Attitude Detection} 
The system uses the wearable STEVAL-STLKT01V1 (shown in Fig.\ref{fig.st}(b)) module to determine the attitude of the target and report it in real-time. The module judges the real-time attitude of the target through an acceleration sensor LSM6DSL. Once there is an abnormal state of acceleration value (such as falling down), the abnormal alarm function of the cloud will be triggered. At the same time, the magnetometer and microphone are used to determine the direction and sound acquisition of the user. By integrating the volume data recorded by the microphone with the compass Angle data, the sensor can record the noise volume in the environment and the geographic orientation of the target. The module will connect to the phone via Bluetooth and then use the smartphone as a gateway to upload data to the cloud in real-time via cellular networks or Wi-Fi.

\paragraph{Environmental Monitor} 
The system will use a temperature and humidity sensor (LPS22HB) to give real-time feedback on the environment of the target location. The data returned by sensors on the B-L475E-IOT01A development board (shown in Fig.\ref{fig.st}(a)) can monitor the environmental variable information of the development board in real-time. The collected data will be directly connected to the core network through the Wi-Fi module (Inventek ISM43362-M3G-L44) that comes with the development board and uploaded to the cloud. The authentication process between the device and the cloud will be done directly on the development board.

\subsection{Data Uploading}
In this section, methods for uploading local data to the cloud and the network protocols involved are described. In this system, two edge networks based on LoRa and Wi-Fi are established. In this paper, the details of the LoRa gateway and the identity authentication between the local device and cloud are mainly introduced.

\paragraph{LoRa Gateway} 
LoRa (Long Range) is a communication protocol and system architecture designed for long-distance LoRa telecommunications networks\cite{4}. In this system, a LoRa gateway is deployed in the edge network where the data collection device is located, so that LoRaWAN is set up to collect data fed back by the sensing devices. The data streams will be uniformly processed at the gateway, such as adding encapsulation and time stamp. The gateway accesses the core network through the Ethernet protocol and sends the processed data to the corresponding interface in the cloud server through the MQTT protocol. The MultiConnect Conduit IP67 base station used in this system is a rugged IoT gateway solution designed for outdoor LoRa public or private edge network deployments. The device supports LoRaWAN applications in almost any environment.

\paragraph{MQTT Broker} 
MQTT (Message Queuing Telemetry Transport) protocol is a commonly used network protocol in the field of Internet of things\cite{5}. In this system, the transformation layer communicates with the cloud connection layer through the core network through this protocol. MQTT is a connection protocol that specifies how data bytes are organized and transferred over a TCP/IP network. MQTT is typically used to build a network of sensors where various sensors can publish sensor values in the form of their sensor-specific messages. Subscribers can subscribe to different messages for appropriate operations. The MQTT broker processes messages that are forwarded from the publisher to the subscriber.

\paragraph{Authentication} 
With millions of sensors around the world, the IoTs contains more and more private information about people's daily behaviours, health and activities, and most of the information is exposed in non-professional scenarios, such as the smart home. With the development of professional technology and resources used by hackers, it is becoming more and more important to ensure the confidentiality, authenticity and integrity of information in the IoTs\cite{3}. 

In this system, asymmetric key encryption is used as the method of establishing a connection between the forwarding layer and the connection layer to ensure the security of data. Anyone with access to the public key can send an encrypted message to the private key holder, and only the private key holder can decrypt and read the message. Also, the public and private keys allow the document to be signed, and the private key is used to add a digital signature to the message. Anyone with a public key can check the signature and know that the original message has not been changed. In addition to indicating that the message has not been tampered with, a digital signature can also be used to prove the ownership of the private key. Anyone with a public key can verify the signature and ensure that the signer has the private key when signing the message.

\subsection{Cloud Server}
In this section, the connection layer data receiving method and the data storage layer principle, as well as the cloud resource management will be introduced. This system adopts cloud computing to process and distribute the data in the system centrally. Using a cloud platform as a transfer station for data exchange and processing, any device connected to the core network can use the public network interface provided by the system and join the created network after being authorized by the system. Users can also receive feedback from the system through the user interface provided by the system at any point accessing to the network. 

\paragraph{Device Shadow} 
All the functions in the system have corresponding device shadow in the cloud. The device shadow is a JSON document that can be viewed as a map of the local device in the cloud and can be used to store and retrieve the current state of the device. The device can publish updated status information into the device's shadow, synchronize status when establishing a connection, or publish its current status information into the shadow for use by a program or other device. The system uses the rules engine to provide data flow and scenario links by configuring SQL-based rule statements to transfer device data to other functions or devices, such as storage and computation.

\paragraph{Database} 
In this system, the table storage model serves as the database for the data storage layer. A table is a collection of items, and each item is a collection of attributes. After the data is received according to the format of the connection layer SQL statement, the timestamp is used as the primary key (row identifier), and the data category is stored in the form through the column identifier. When the front end reads the data, it can read the current real-time data according to the timestamp and show the trend of the data over a period of time. To avoid large amounts of outdated data redundancy taking up database space and potential problems caused by old data, most of the data in the form is currently set to be deleted automatically one day after expiration.

\paragraph{Access Management} 
In a complex system, to ensure the security and stability of cloud resources, it is necessary to prevent users of different jobs and software from modifying functions irrelevant to them. Therefore, it is necessary to reduce information security risks by assigning as few permissions as possible to roles in the project. In this system, different permissions are set for database writing, front-end data reading, etc., to ensure that the functions only run within the scope.

\subsection{User Interface}
The system provides a real-time user interface in the service layer to ensure that users can get the latest information on the system targets anytime and anywhere. This system provides two access methods: 1) directly access the web version interface through the browser, and 2) access the mobile version user interface through the mobile APP. Additionally, for users with mobility or dyslexia, the system provides 3) a voice broadcast device that provides voice feedback in real-time.

\begin{figure}[htbp]
\centering{\includegraphics[width = 80 mm]{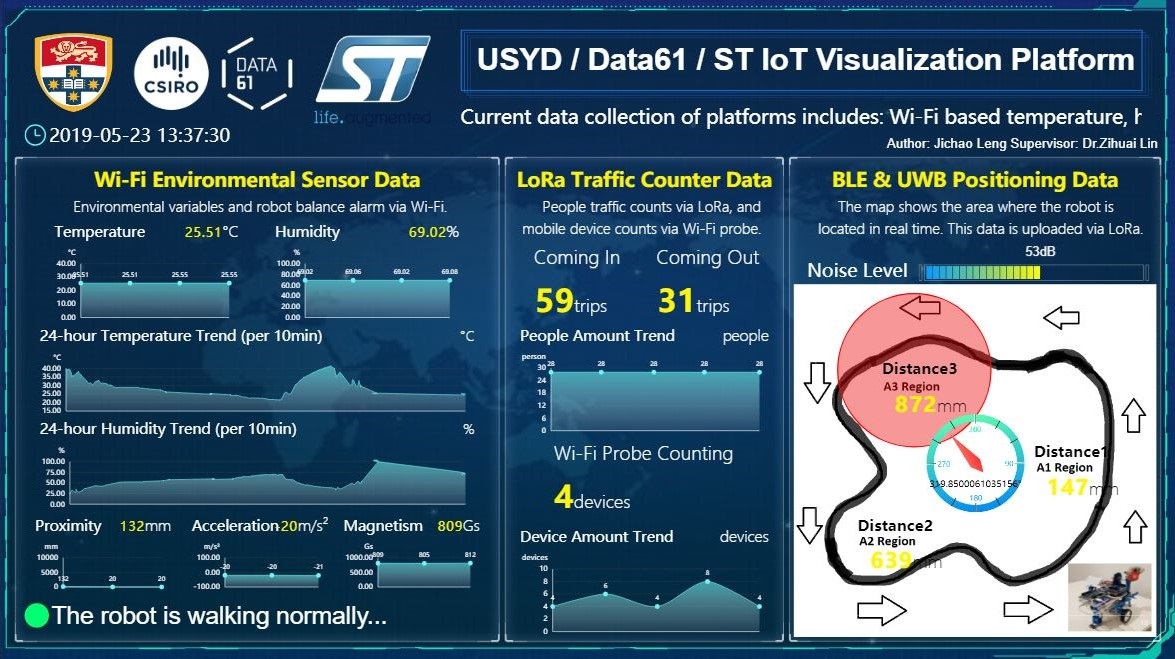}}
\caption{Web Interface.}
\label{fig.web}
\end{figure}

\paragraph{Web Interface} 
In this system, the visual interface takes the corresponding form in the database as the data source and presents the data analysis results to the user in a graphical form. Because a timestamp is added to each message during the data store, the current latest data can be read by finding the data row that contains the maximum timestamp value, and the message from the corresponding function can be read by selecting the datatype identity. The whole visual interface is divided into three parts: the first part is real-time monitoring of environment variables, the second part is people traffic statistics, and the third part is target positioning. The interface is shown in Fig.\ref{fig.web}.

\begin{figure}[htbp]
\centering{\includegraphics[width = 50 mm]{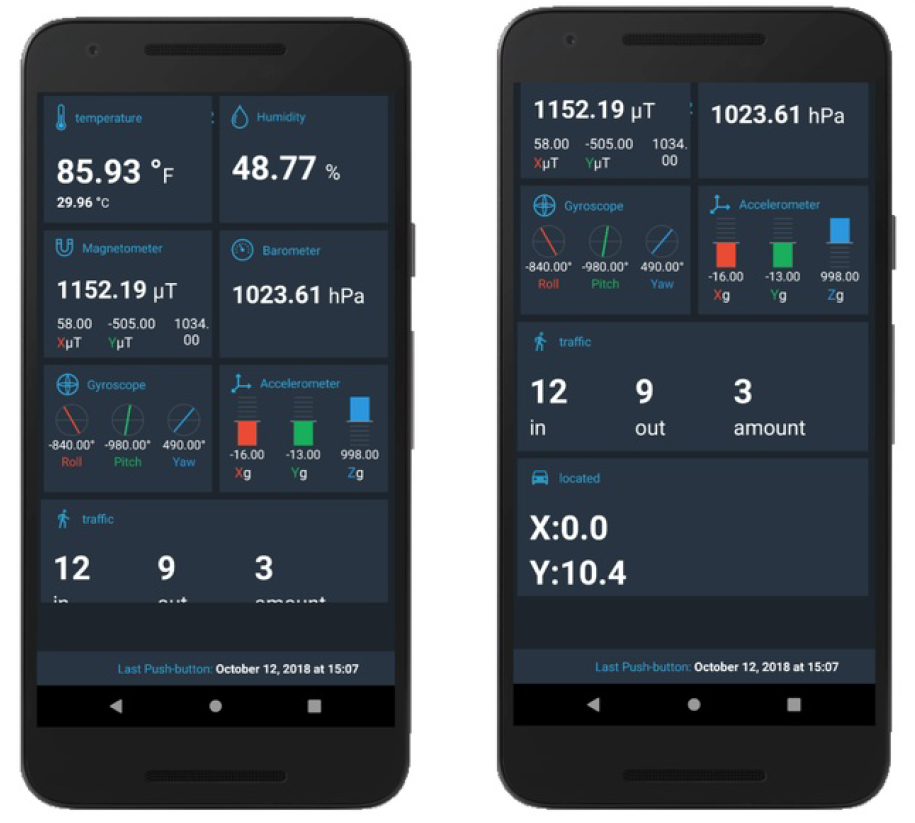}}
\caption{Mobile Interface.}
\label{fig.app}
\end{figure}

\paragraph{Mobile Interface} 
In order to facilitate users to get information anytime and anywhere, a mobile APP is developed for this system to allow users to access the user interface directly using the mobile phone. The application reads the data from the corresponding database through the HTTP protocol and presents it to the user through the graphical interface. The mobile interface is shown in Fig.\ref{fig.app}.

\paragraph{Voice Broadcasting} 
The voice broadcast system uses the STM32F769 device, which can subscribe to the MQTT topics to ask for the messages received by the cloud from the function under the Wi-Fi connection, and convert the text messages into voice broadcast to the user by XFS5152.

\section{Demonstration}
\begin{figure}[htbp]
\centering{\includegraphics[width = 80 mm]{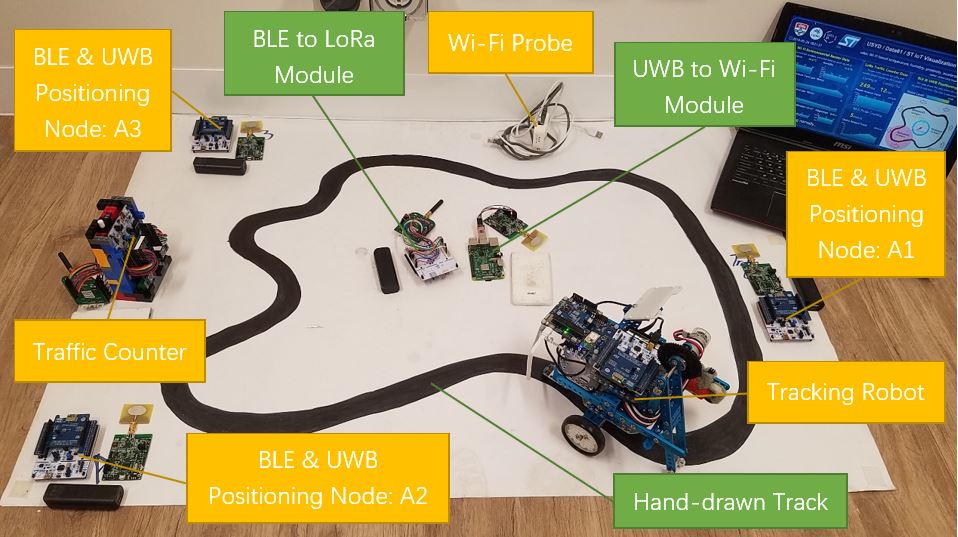}}
\caption{Demo Site.}
\label{fig.demo}
\end{figure}

To verify the system, we design a demonstration to test the performance of some of the functions in our system. The system uses an automatic tracking robot to demonstrate existing functions. The project demo site in Fig.\ref{fig.demo} uses the map of the pre-set track, and the robot will move along the track on the map. 

\begin{figure}[htbp]
\centering
\begin{tabular}{cc}
\includegraphics[width = 68 mm]{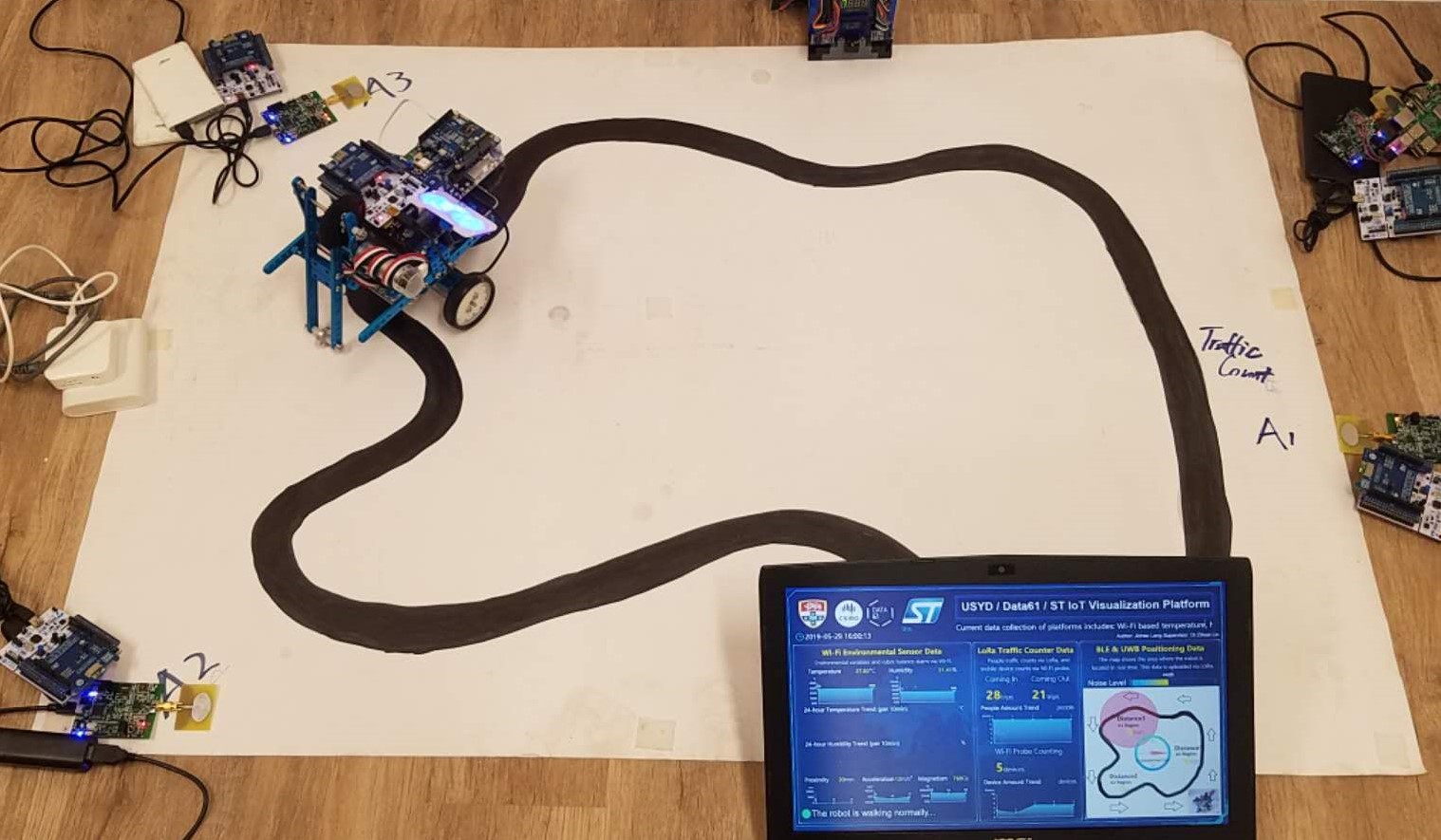}
\\
(a) Indoor Positioning.
\\
\includegraphics[width = 68 mm]{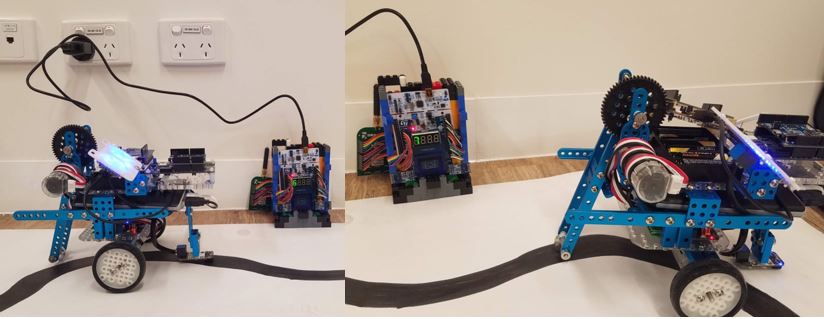}
\\
(b) Flow Statistics.
\\
\includegraphics[width = 68 mm]{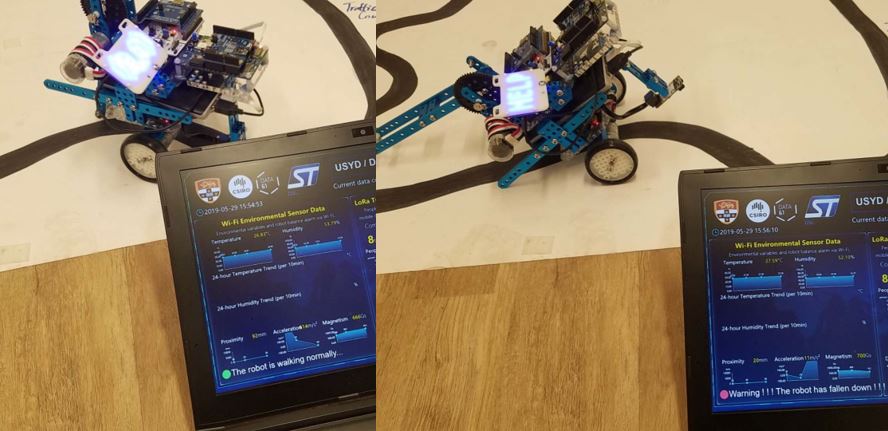}
\\
(c) Fall Alarm.
\\
\end{tabular}
\caption{Functions Demo.}
\label{fig.func}
\end{figure}

In Fig.\ref{fig.func}(a), three BLE anchor nodes and UWB anchor nodes are placed around the map. When the robot approaches any node during the journey, the system can determine the current location of the robot. In Fig.\ref{fig.func}(b), the people traffic counting module is placed on the map to increase the count when the robot passes through the counting position, and the results will be recorded in the cloud. In Fig.\ref{fig.func}(c), during this process, the robot periodically flips its arm to demonstrate a fall alert. At the same time, the environment monitoring sensor carried by the robot will reflect the changes in the environment at any time.

\begin{figure}[htbp]
\centering{\includegraphics[width = 55 mm]{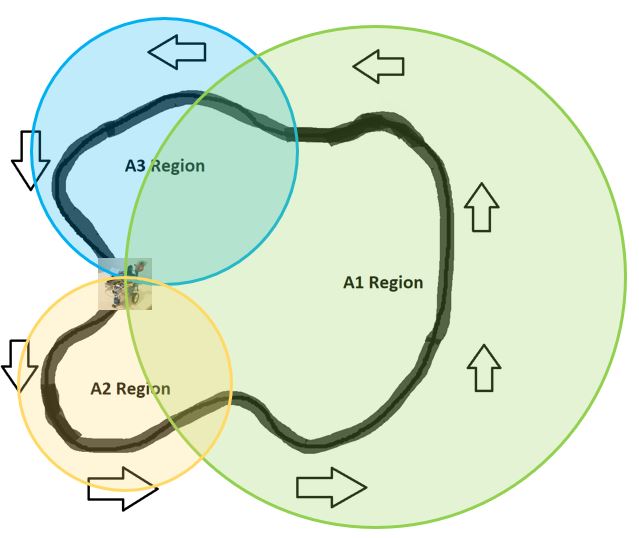}}
\caption{UWB Trilateral Positioning.}
\label{fig.tril}
\end{figure}

For the indoor positioning function, the three anchor nodes are distributed in the three corners of the map. The system uses the triangulation method to confirm the real-time position of the robot and uploads the position to the cloud, which is fed back in real-time by the visual interface. The schematic diagram of the positioning method is shown in Fig.\ref{fig.tril}. 

To find the ranging accuracy of this indoor positioning function, we measure the actual distance between the UWB anchor point and the target with a laser rangefinder and compared it with the distance value fed back by the UWB positioning function. The results are shown in Table.\ref{tab.dis}. According to the test results, the accuracy of UWB positioning can reach the decimeter level.

\begin{table}[htbp]
\caption{UWB Ranging Accuracy Test}
\centering
\begin{tabular}{ccc}
\hline
Actual Distance (m)	&	Measure Results (m)	\\
\hline
1.325	&	1.17\\
2.257	&	2.1\\
2.665	&	2.5\\
3.227	&	3.08\\
4.213	&	4.1\\
\hline
\end{tabular}
\label{tab.dis}
\end{table}

\begin{figure}[htbp]
\centering{\includegraphics[width = 87 mm]{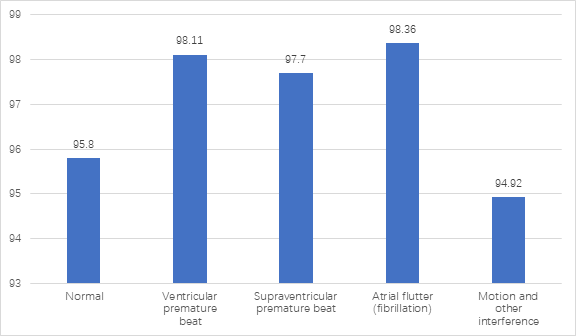}}
\caption{ECG Model Accuracy.}
\label{fig.ecgacc}
\end{figure}

For ECG signal detection, a set of ECG samples is used to verify the accuracy of the model by comparing the judgment results of the system model with the expert's diagnosis. For different types of heart disease, the calculation method of mean average precision (mAP) is shown as follows, and the results is shown in Fig.\ref{fig.ecgacc}, where AP stands for average precision. For this case, the calculated result is 96.978\%.
\begin{equation*}
mAP=\frac{1}{\left | Q_R \right |}\sum_{q\in Q_R}{AP(q)}
\end{equation*}

\section{Conclusion}
In this paper, a set of intelligent hospital IoT system is introduced, which provides indoor positioning, ECG and attitude detection, people flow statistics, environmental monitor functions, and it also provides the various real-time user interface for users to obtain data feedback. For healthcare scenarios, authenticity and reliability of network connections and confidentiality of data are also important issues in the Internet of things. Once the system is attacked and paralyzed, the consequences for the patient can be devastating. Similarly, once the data in the database is leaked or even tampered with, the most serious consequence may be that doctors misjudge the wrong information in the diagnosis process, causing worse results. In this system, asymmetric key encryption technology, which is relatively common and relatively secure, is adopted between the device and the cloud server to realize two-way authentication. This approach can avoid some potential message interception attacks and man-in-the-middle attacks.

In future work, this system will attempt to integrate the functional equipment of edge network with SDN to improve the efficiency and stability of the network. At the same time, the system will add more functions, such as EEG and other professional medical monitoring devices, to integrate professional medical field data into the system. During the data upload phase, the system will add more network access. Diversified network technical support enables the system to select the most appropriate network protocol according to specific deployment conditions of different scenarios, thus improving efficiency and reducing deployment and maintenance costs\cite{6}. We will also develop novel algorithms to improve the network reliability based on our existing works on wireless communications, e.g.,  
\cite{b3a,b3b,b3c,distributedRaptor,Raptor_ML,JNCC,RCRC,codedcpm1,codedcpm2,codedcpm3,NC1,NC2,NC3,NC4,WRN,cellular1,cellular2,cellular3,MIMO_capacity,network_capacity, UAVdownlink,UAV_THz,UAV_2}. 
Privacy issues  \cite{privacy1,privacy2,privacy3,privacy4} will also be our future work.

\end{document}